# Modeling Mechanical Properties of Aluminum Composite Produced Using Stir Casting Method

MUHAMMAD HAYAT JOKHIO*, MUHAMMAD IBRAHIM PANHWAR**, AND MUKHTIAR ALI UNAR***

## ABSTRACT

ANN (Artificial Neural Networks) modeling methodology was adopted for predicting mechanical properties of aluminum cast composite materials. For this purpose aluminum alloy were developed using conventional foundry method.

The composite materials have complex nature which posses the nonlinear relationship among heat treatment, processing parameters, and composition and affects their mechanical properties. These nonlinear relation ships with properties can more efficiently be modeled by ANNs. Neural networks modeling needs sufficient data base consisting of mechanical properties, chemical composition and processing parameters. Such data base is not available for modeling.

Therefore, a large range of experimental work was carried out for the development of aluminum composite materials. Alloys containing Cu, Mg and Zn as matrix were reinforced with 1- 15% $Al_2O_3$ particles using stir casting method. Alloys composites were cast in a metal mold. More than eighty standard samples were prepared for tensile tests. Sixty samples were given solution treatments at 580°C for half an hour and tempered at 120°C for 24 hours.

The samples were characterized to investigate mechanical properties using Scanning Electron Microscope, X-Ray Spectrometer, Optical Metallurgical Microscope, Vickers Hardness, Universal Testing Machine and Abrasive Wear Testing Machine.

A MLP (Multilayer Perceptron) feedforward was developed and used for modeling purpose. Training, testing and validation of the model were carried out using back propagation learning algorithm.

The modeling results show that an architecture of 14 inputs with 9 hidden neurons and 4 outputs which includes the tensile strength, elongation, hardness and abrasive wear resistance gives reasonably accurate results with an error within the range of 2-7 % in training, testing and validation.

Key Words: Neural Network Modeling, Mechanical Properties, Aluminum Cast Composite

## 1. INTRODUCTION

Neural networks have been applied in every area for solving various materials related problems such as design and development, processing and controlling of materials, products and equipment. This technique has also been effectively applied by many of investigators in the field of materials

* Associate Professor, Department of Metallurgy & Materials Engineering, Mehran University of Engineering and Technology, Jamshoro.
** Professor, Department of Mechanical Engineering, Mehran University of Engineering and Technology, Jamshoro.
*** Professor, Department of Computer Systems Engineering, Mehran University of Engineering and Technology, Jamshoro.

science and engineering [1-23].

The applications of neural networks modeling was adopted by Badeshah [1], Shah, I., [2], and Genel, et. al. [3] who had well documented neural networks applications in materials science concerned with the microstructural evaluation of steels, processing and properties of steels as well as conducted study on ductile cast iron. The materials science based work using neural networks which had brought attention of researchers recently conducted by Sha and Edwards [4].

Microstructural features such as amount of austenite retained in ductile austempered cast iron was estimated by Yascas, et. al. [5] using artificial neural networks. The features classification of alloy steel microstructures consist ferrite and pearlite was investigated using back propagation algorithm which can effectively be used for the features classification by Martin and William [6]. The microstructure image analysis of complex systems have been determined by using neural networks technique as reported by Maly, Harck and Novotny [7].

In the field of powder metallurgy neural networks had been successfully applied by Sudhakar and Hague [8], Ohdar and Pasha [9], and Jokhio, M.H., et. al. [10] in determining the mechanical properties and processing parameters.

Many of investigators such as Kowalski, et. al. [11], Dobrzanski and Sitech [12] applied neural networks modeling in processing of steels and determining their performance as well as characterization involving the complex analysis of problems. The materials performance depends on the complex materials, interrelated factors including chemistry and processing method whereas, experimental observations could not capture all materials aspects. Material development, processing and characterizations are timed consuming and difficult tasks. However, neural networks have the capability in capturing the experimentally observed behavior through a learning process Jokhio, M.H., [13-14].

## 2. EXPERIMENTAL PROCEDURE

Modeling mechanical properties of aluminum cast composites materials needs sufficient data. However, some information is available on modeling mechanical properties of composite materials processed via conventional stir casting method, which is insufficient for neural networks modeling. Therefore, a comprehensive experimental work was conducted for data generations required for modeling mechanical properties.

### 2.1 Strategy for Alloy Development

For this purpose pure aluminum, magnesium, zinc, copper metals as ingots and $Al_2O_3$ as powder were purchased. Six master alloys were prepared through conventional foundry method. The detailed compositions in weight percentage in grams are given in Table 1.

For manufacturing of alloys pure aluminum was melted in a pit furnace using graphite crucible at Mehran University of Engineering & Technology, Workshop as shown in Fig. 1.

TABLE 1. SELECTED COMPOSITIONS OF MASTER ALLOYS IN GRAMS FOR MANUFACTURING OF COMPOSITE MATERIALS

| Alloy No. | Cu (gm) | Mg (gm) | Zn (gm) | Al (gm) |
|-----------|---------|---------|---------|---------|
| 1. | 0.00 | 122 | 150 | 4728 |
| 2. | 100 | 140 | 300 | 4460 |
| 3. | 400 | 50 | 300 | 4250 |
| 4. | 250 | 50 | 0.00 | 4700 |
| 5. | 150 | 120 | 300 | 4430 |
| 6. | 200 | 125 | 300 | 4375 |

The various alloying elements such as Cu, Zn and Mg were added and mixed manually for 10 minutes. The prepared molten alloys were cast in sand molds.

## 2.2 Manufacturing and Casting of Composite Samples

Each alloy was remelted in an electrically heated bath Fig. 2. The alloy weighted quantity 500gm was melted at $850^{\circ}C$ for 30 minutes Fig. 3. $Al_2O_3$ particles in weight percentage of 2.5, 5, 10, and 15% were preheated and mixed in aluminum matrix for 10 minutes shown in Fig. 4. The alloys were purged for few second with nitrogen gas Fig. 5. The prepared composite material was cast in a steel mold. The size of the casted steel mold is 1/2 inch dia and

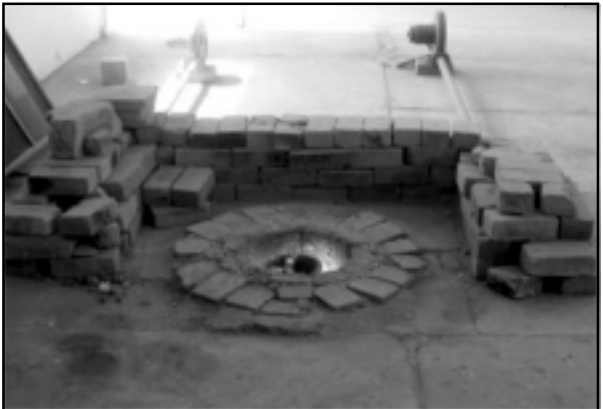

FIG. 1. GAS FIRE FURNACE

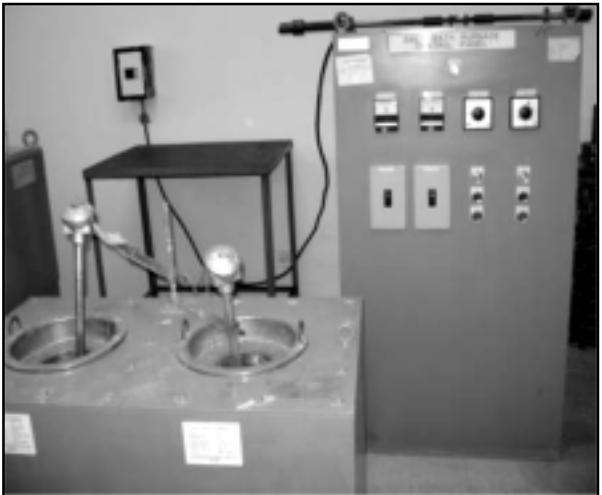

FIG. 2. ELECTRICALLY HEATED BATH

8 inches length which were drilled on opposite sides for evolutions of gases in order to minimize the porosity contents as shown in Fig. 6.

## 2.3 Samples Preparation

100 standard specimens were prepared at Mehran University of Engineering & Technology, workshop using lath machine for tensile tests (Figs. 7-8). To avoid heating of samples coolant was used during the machining. The standard specimen size consists of 11.28mm dia, 56 gauge length, radius r=6mm and 130mm total length.

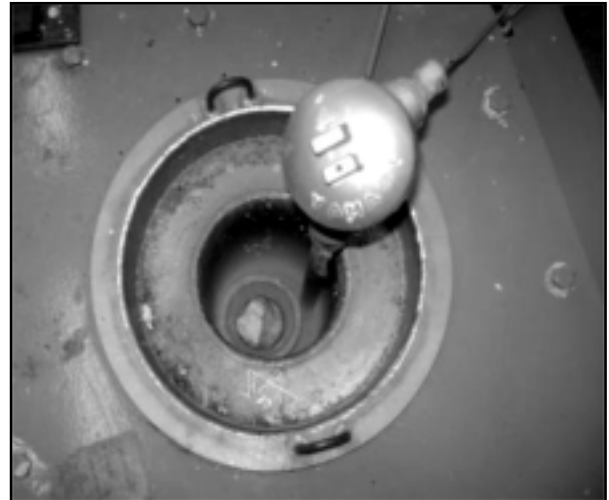

FIG. 3. PREPARATION OF MASTER ALLOY

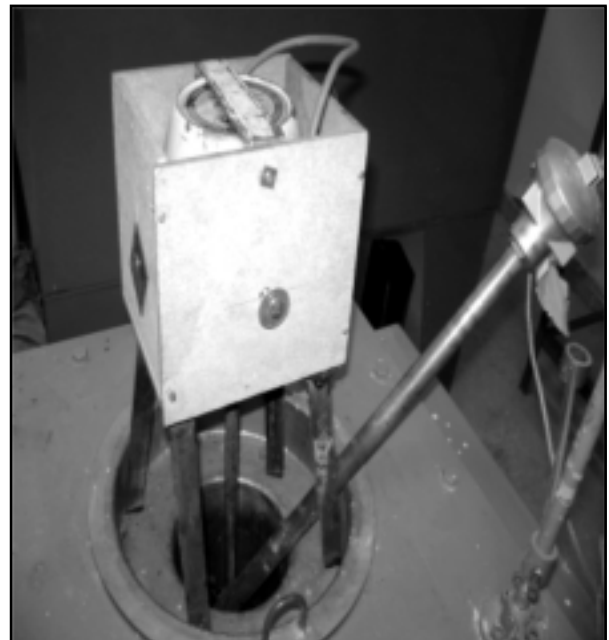

FIG. 4. MIXER ARRANGEMENTS

## 3.4    Precipitation Treatment

Sixty (60) specimens were heat treated using electrically heated muffle furnace. Specimens were heated at 580ºC

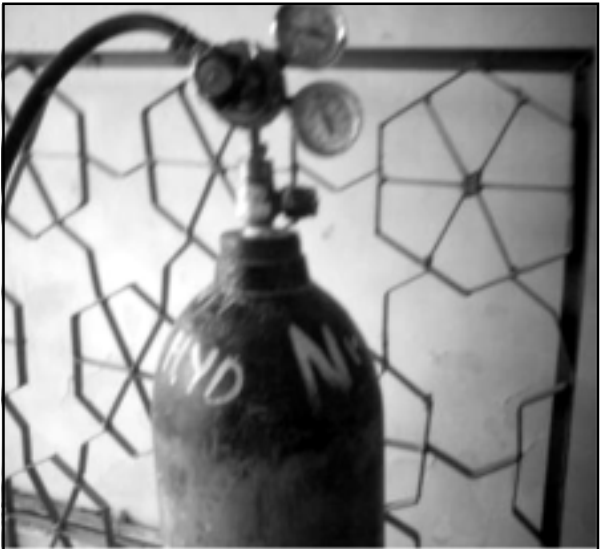

*FIG. 5. N$_2$ GAS CYLINDER*

for half an hour and quenched in water (solution heat treatment). All specimens after solution treatment were tempered at 120ºC for 24 hours.

## 2.5    Tensile Test

Universal Testing Machine was used for tensile tests of all samples. Tensile tests were conducted at the Department of Metallurgical Engineering, Dawood College of Engineering & Technology, Karachi, Pakistan. The tensile test parameters are given in Table 2.

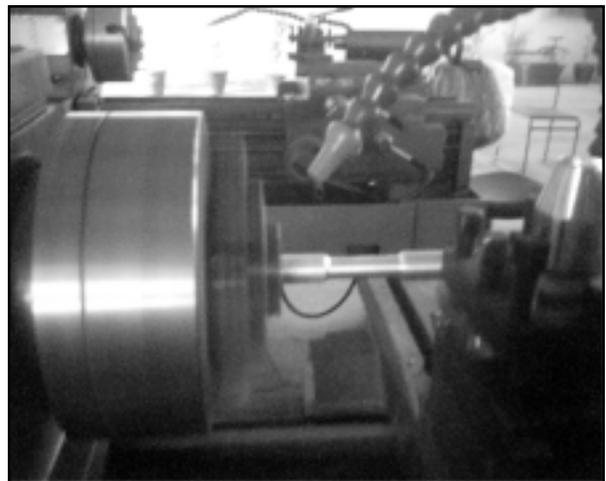

*FIG. 7. MACHINING OF SAMPLES ON LATH*

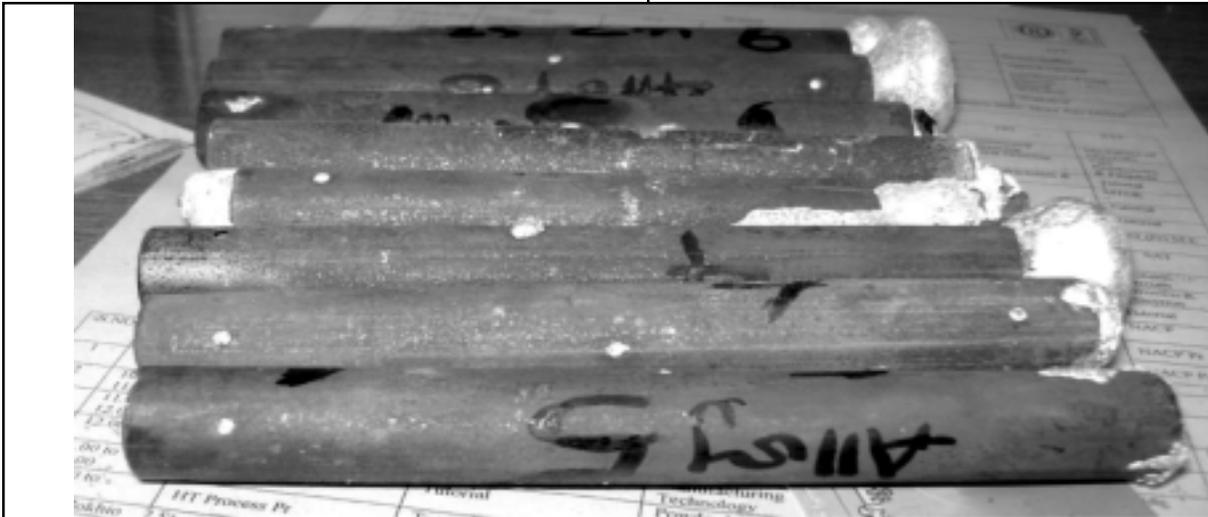

*FIG. 6. PREPARED SAMPLES IN METAL MOLD*

## 2.6 Chemical Analysis

Samples size 15mm dia and 15mm length were prepared for chemical analysis of specimen using Spectrometer available at Dawood College of Engineering & Technology, Karachi and Scanning Electron Microscope which is available at Mining, Engineering Department Mehran University of Engineering and Technology Jamshoro (Fig. 9). Aluminum oxide particles size was also analyzed using Horiba particle size analyzer (Fig. 10).

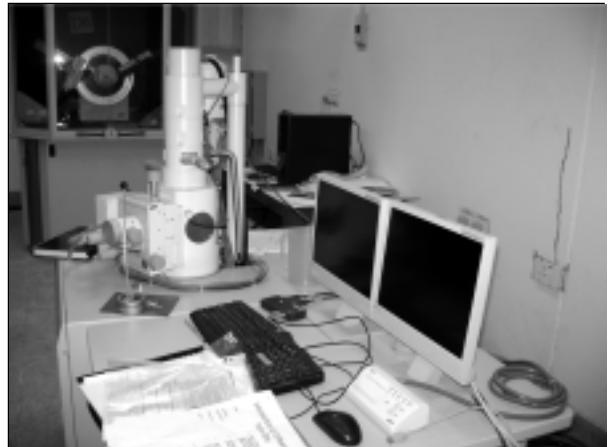

*FIG. 9. SEM USED FOR ANALYSIS*

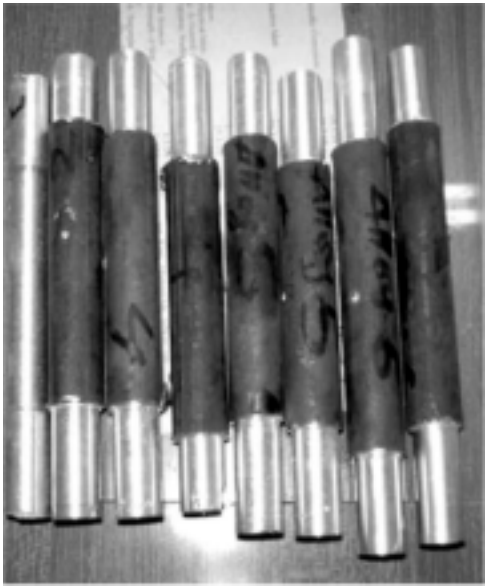
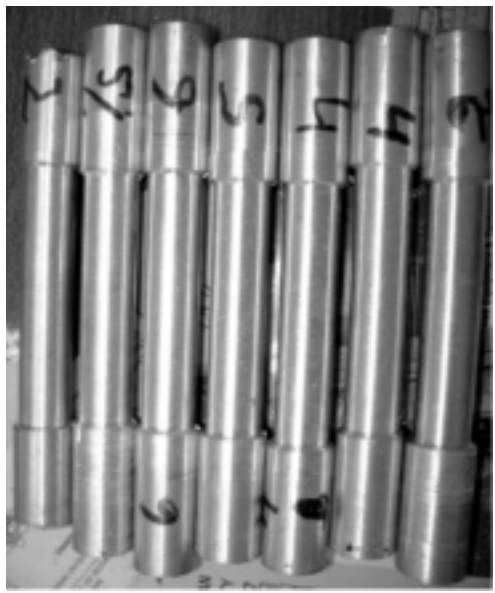

*FIG. 8. STANDARD SPECIMENS PREPARED FOR TENSILE TEST*

**TABLE 2. PRAMETERS OF TENSILE TESTING MACHINE**

| | |
|---|---|
| Order Number | 12.5  Cast Sample |
| Charge | Rs 300 |
| Test Standard | ASTM |
| Customer | M. H JOKHIO |
| Material | Aluminium Cast Composte Material |
| Extensometer | |
| Load Cell | Maximum 250 KN |
| Pre-Load | 2  N/mm² |
| Pre-Load Speed | 10 mm/min |
| Test Speed | 50 mm/min |

## 2.7 Hardness Test

Vickers hardness Testing Machine at 1 kg load was used for hardness test. Four readings of each sample were taken and mean diagonal area was calculated for recording VHN value.

## 2.8 Abrasive Wear Test

Pin on disc method was used to determine the abrasive wear of samples. All samples were weighted on electronic balance (Fig. 11) and samples were fixed in machine attachment for abrasive wear test (Fig. 12). Figs. 11-13 shows the over all setup for abrasive wear test. The test was carried out using 150 mesh emery paper fixed on disc with steel frame for abrasive wear test.

This paper was removed after each 15 minutes. Four readings of each sample were taken for wear test. Whereas the machine speed was fixed at 1500 rpm; at a load of 1 kg.

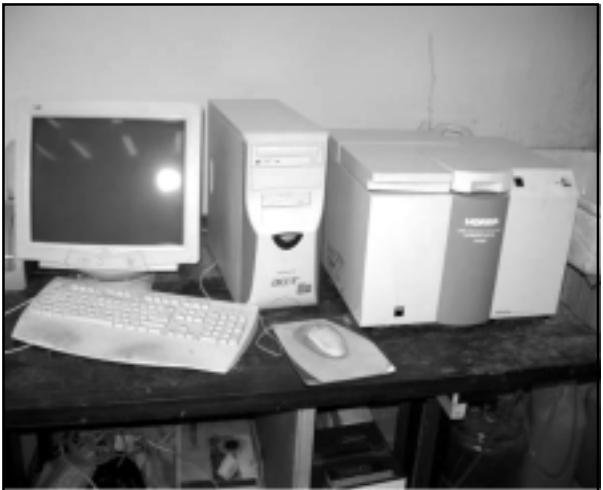

*FIG. 10. HORIBA PARTICLE SIZE ANALYZER*

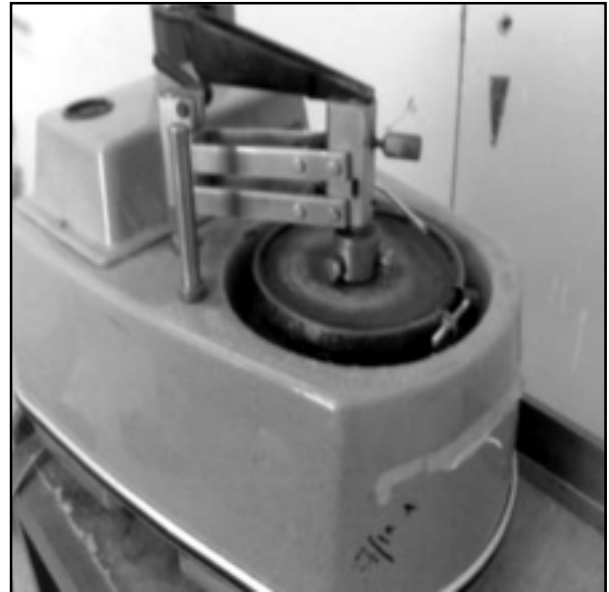

*FIG. 12. WEAR TESTING MACHINE*

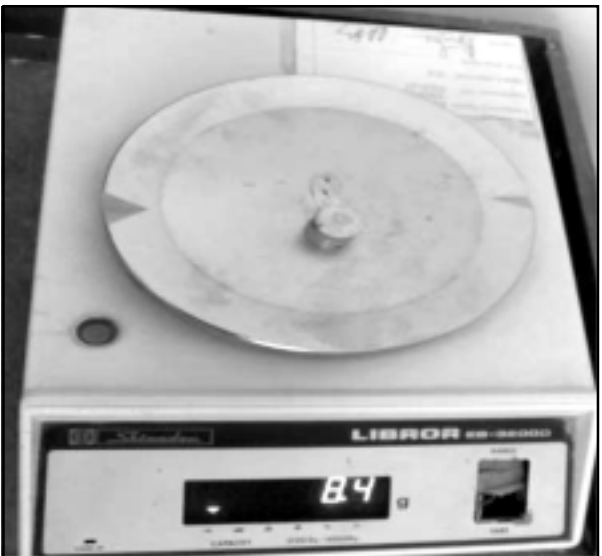

*FIG. 11. ELECTRONIC BALANCE*

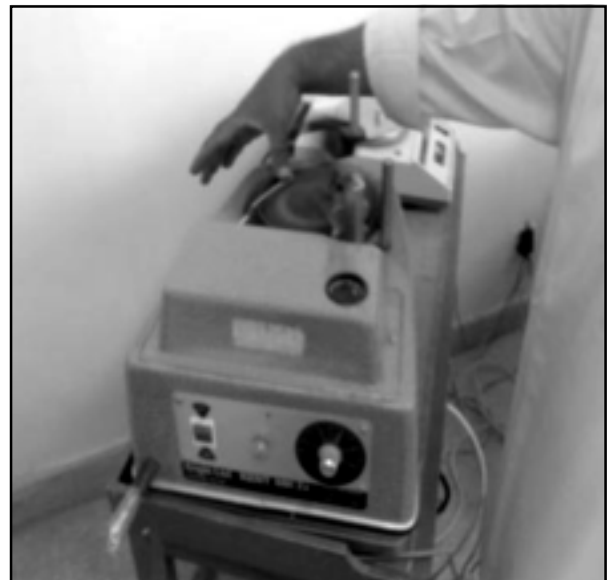

*FIG. 13. ABRASIVE WEAR SETUP.*

Water was used as lubricants with a constant flow rate. Total weight loss in grams and average weight loss in grams were calculated after taking four readings of each sample.

## 2.9    Measurement of Density and Porosity

Densities of all specimens were measured through weight in air divided by volume method. The porosity of samples was calculated through the theoretical density of specimens which were calculated from the average composition from the elemental density of metals as given in Tables 3-4.

## 2.10    Metallography

Optical Metallurgical Microscope at Mehran University of Engineering & Technology, Jamshoro, and Pakistan Steel Mills Laboratories were used for metallographic studies. Scanning Electron Microscope (Fig. 9) was also used for few samples. The samples were metallographically prepared after grinding at 120-1200 mesh emery papers and samples were polished using alfa aluminum oxide powder. Samples were etched in Keller's solution.

## 2.11    Modeling Methodology

The modeling methodology for predicting the mechanical properties of cast composite depends on the nature of problem. The present researches involves the prediction of mechanical properties of aluminum alloy base casting composite materials. The performance composite materials depend on chemical composition, processing conditions. The composition, processing parameters and heat treatment variables have strong influence on properties. Obviously complicated nonlinear relationship exists between chemistry, processing conditions and their resulting properties. To search for optimum composition for the desired properties a nonlinear relationship is to be established for good results. This requires the selection of input and output parameters as well as the architecture of the networks and their learning algorithm.

Literature reveals that the complexity of this nonlinear behavior of materials and related performance can more efficiently be modeled using ANNs.

Unar, M.A., [15] Talpur, M.I.H., [16], Jokhio, M.H., et al [9] have referred that the artificial neural network modeling is a comparatively new field of artificial intelligence which tries to mimic the structure and operation based on biological neural systems such as human brain by creating an artificial neural network on computers.

Therefore, ANN modeling methodology is used for modeling mechanical properties of aluminum cast composite material. The selection of inputs layer, hidden layer, outputs layer as well as the selection of learning algorithm, training and testing (validations) of model is required for efficient neural network modeling as described below:

**TABLE 3. AVERAGE COMPOSITIONS USED FOR CALCULATION OF THEORETICAL DENSITY OF ALLOYS**

| Alloy No. | Si | Fe (%) | Cu (%) | Mg (%) | Zn (%) | Al (%) | Theoretical Density (gm/cm³) |
|---|---|---|---|---|---|---|---|
| 1. | 1.1 | 0.92 | 0.066 | 2.21 | 3.0 | 92 | 2.85 |
| 2. | 2.8 | 0.72 | 2.45 | 2.15 | 5.0 | 84 | 2.99 |
| 3. | 3.8 | 0.78 | 7.0 | 1.28 | 5.16 | 75 | 3.17 |
| 4. | 2.6 | 1.0 | 4.7 | 0.6 | 0.6 | 89 | 3.0 |
| 5. | 3.2 | 0.46 | 2.7 | 1.8 | 4.4 | 85 | 2.98 |
| 6. | 4.3 | 0.62 | 3.0 | 2.0 | 5.16 | 81 | 2.99 |

**TABLE 4. THEORETICAL DENSITIES OF METALS**

| Si (gm/cm³) | Fe (gm/cm³) | Cu (gm/cm³) | Mg (gm/cm³) | Zn (gm/cm³) | Al (gm/cm³) |
|---|---|---|---|---|---|
| 2.329 | 7.6 | 8.69 | 1.738 | 7.1338 | 2.7 |

### 2.11.1 Architecture of the Model

The design architecture of the model consist input layer, hidden layer and output layer as shown in Fig. 14. It is a feedforward type of model. In data processing each neuron in layer receives inputs from all the neurons of preceding layer.

In present model the chemical compositions such as Fe, Si, Cu, Mg, Zn, $Al_2O_3$ and Al in weight % age, the processing parameters including the melting temperature in °C, melting time in minutes, tempering temperature in °C, tempering times in minutes and density of the materials in $g/cm^3$ were chosen as input parameters. Mechanical properties such as tensile strength in MPa, elongation in %age, hardness in HV and abrasive wear resistance (average weight loss in grams) were selected as output parameters.

The tangent hyperbolic and linear activation functions were selected as transfer functions in hidden layer and output layer neurons respectively. Model consists of fourteen input neurons, nine hidden neurons and four output neurons.

### 2.11.2 Training of Neural Network Model

Error-back propagation learning algorithm was used for training, testing and validation of model Hayken [17]. During training the network runs repeatedly with different neurons, until the output is satisfactorily accurate. For training of neural network pack propagation algorithm was programmed in MATLAB.

Training is the act of continuously adjusting their connection weights until they reach unique values that allow the network to produce outputs that are close enough to the desired outputs as reported by Jokhio, M.H., [14].

The accuracy of developed model therefore, depends on these weights. Once optimum weights are reached, the weights and biases value encode the networks state of knowledge. Therefore, using network on new case is merely a matter of simple mathematical manipulation of these values [15]

### 2.11.3 Validation of Model

Ten alloys from experimental data base were randomly selected for validation of the model For validation purpose five (5) alloys as in cast condition and five (5) alloys as in heat treated conditions were selected for validation of the trained models for both conditions. The results were compared with the experimentally determined mechanical properties which show good matching of experimental and predicted properties of cast composite materials.

## 3. RESULTS AND DISCUSSION

## 3.1 Modeling and Training

The modeling mechanical properties of aluminum alloy based casting composite material for engineering application requires the huge data base considering nonlinear behavior of various metallurgical factors, which include the processing parameters, composition and heat treatment operations with the desired properties. In present investigations the modeling results consist of the training, testing and validation of the trained model.

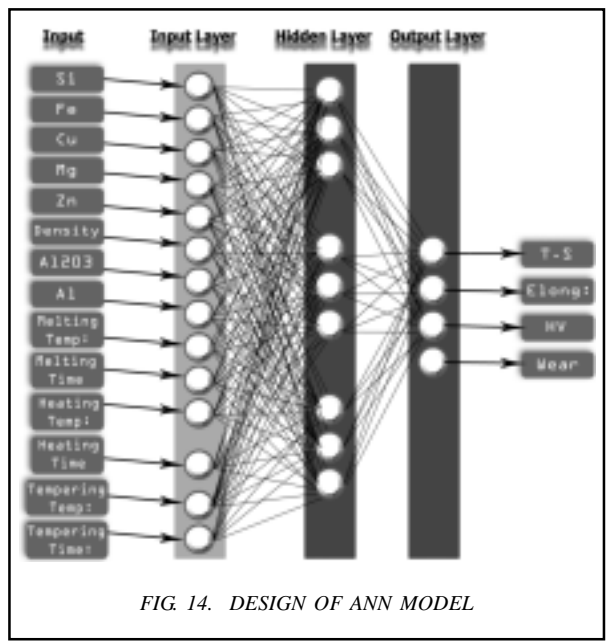

*FIG. 14. DESIGN OF ANN MODEL*

It has been reported in literature that the ANN modeling is powerful tool for modeling, prediction and optimization of properties of materials [Bhadesha, et. al. [1], Li, et. al. [18], and Dobrzanski, et. al. [12]. Sha and Edwards [4] has reported that the performance of artificial neural network modeling depends on model, its database and training algorithm.

Literature reveals that the neural network modeling is, in general a method of regression analysis in which a very flexible nonlinear function such as heat treatment process parameters are to be fitted to the experimental data due to presence of hidden neurons between input and output [15].

The results show that a neuro model containing one hidden neuron is not sufficient to capture the non-linear information from the database [14]. Whereas an increase in the number of hidden neurons can improve the performance of the networks model and the number of hidden layers is also a crucial decision. However, it has been proved mathematically that a single hidden layer feedforward neural network is sufficient for successful results, if enough neurons have been used as reported by Hyken [17]. The important aspect of modeling is the selection of input parameters which affects the output (mechanical properties). On the other hand the selection of a good learning algorithm is also important [19].

This work uses the MLP architecture of feed forward neural networks which were introduced in the 1980s. Since their invention these networks have been used in almost every branch of science and engineering and can be well trained by using error back- propagation algorithm [17].

Despite being 20 years old algorithm its usefulness has never gone down. Therefore, this algorithm dominates in the literature in research of the applied neural networks especially in the field of materials science and engineering. For example Hassan, et. al. [19] have investigated hardness, density, and porosity of aluminum composite by using a MLP network. In modeling they had used 10 neurons in the hidden layer with two inputs and three outputs of the network. A similar work concerned with the effect of heat treatment and abrasive wear resistance was modeled and investigated by Xing, et. at. [20]. Whereas a model structure with 4:10:5, topology was used to predict mechanical properties of hot rolled carbon steel bar by Ozerdum and Kolukisa [21].

Looking at the success of present design MLP network in other applications, it has been decided to use this architecture of networks for predicting mechanical properties of cast composites under investigation. The model parameters are as already mentioned in Fig. 14. During the training of model about 80 data sets were prepared and fed for training. After extensive simulation works the results were found satisfactory with nine neurons in the hidden layer were found to be sufficient for this application. The SSE (Sum of Square Error) various numbers of epochs is plotted in Fig. 15. The SSE of 0.0001 was achieved only after 70 epochs within few seconds which shows the model is well trained with sufficient accuracy with adaptive learning rate and momentum.

Considering the training results of mechanical properties including the tensile strength, elongation, hardness and abrasive wear as shown in Figs. 16-19. The training results show these mechanical properties such as tensile strength and hardness data were well trained as compared to elongation However, the trained data is within acceptable range of the variation 2-7% error as shown in Figs. 16-19.

The SSE, SSW and effective number of parameters of the model were determined to be 0.24, 124.7 and 12.19 respectively as shown in Fig. 20.

Therefore, using trained models knowledge on new case is merely a matter of simple mathematically manipulation of these values which will also be used to determine the validity of the model.

For the testing of validity of the generalization performance of the present trained model with the actual experimental results of the unseen data which were feeding to the model is essential for validation of any model. For this purpose 10 datasets, five from as cast and 5 from as heat treated conditions were used for validity measurement. The inherent knowledge spectrum of the trained network give results which were compared with the experimental results are given in Table 5 and shown in Figs. 21-24. The mechanical properties such as tensile strength elongation, hardness and abrasive wear resistance as determined by the model are very close to the experimentally determined mechanical properties which show the validity of model.

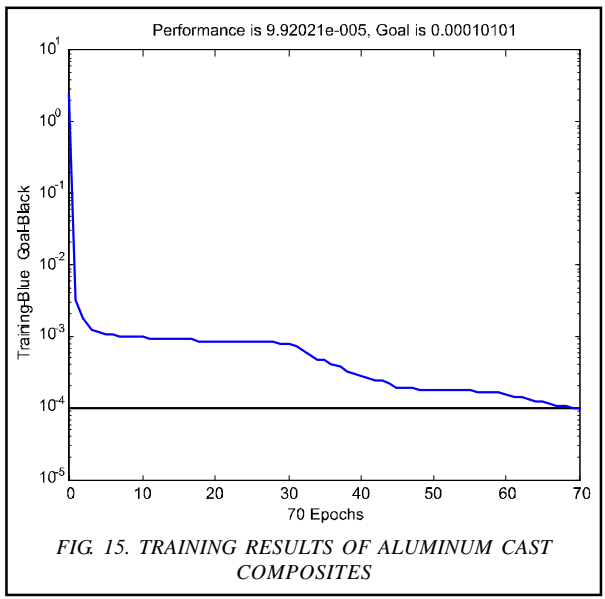

FIG. 15. TRAINING RESULTS OF ALUMINUM CAST COMPOSITES

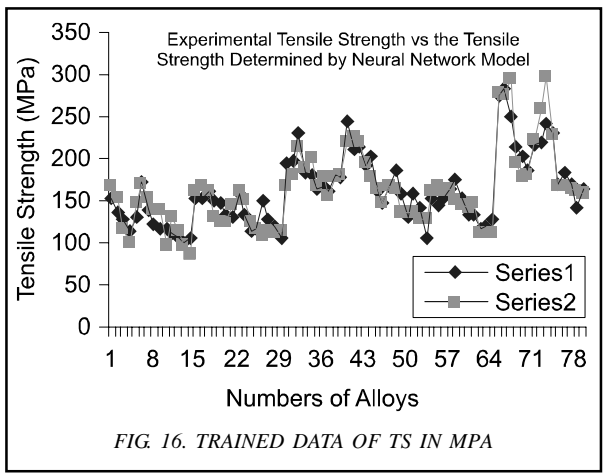

FIG. 16. TRAINED DATA OF TS IN MPA

Some of the previous work concerned with modeling mechanical properties was conducted by Hassan, et. al. [19]. They investigated mechanical properties of compocast aluminum composite reinforced with SiC. They had selected 46 data set for training and only 8 data for validation. A comprehensive research work was conducted recently by Altinkok and Koker [22-23]. They had predicted tensile strength and density in particles reinforced in aluminum composite using neural network modeling and they used 15 datasets for training and 5 dataset for validation.

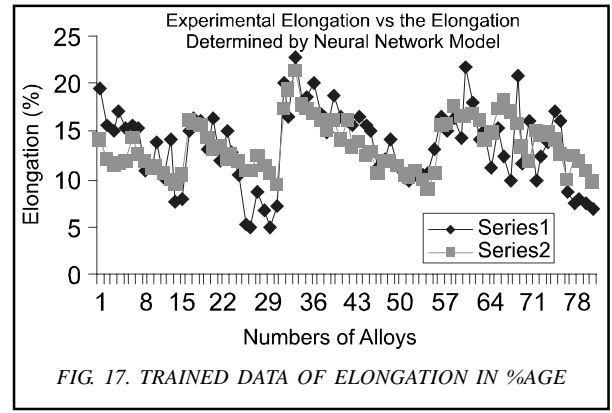

FIG. 17. TRAINED DATA OF ELONGATION IN %AGE

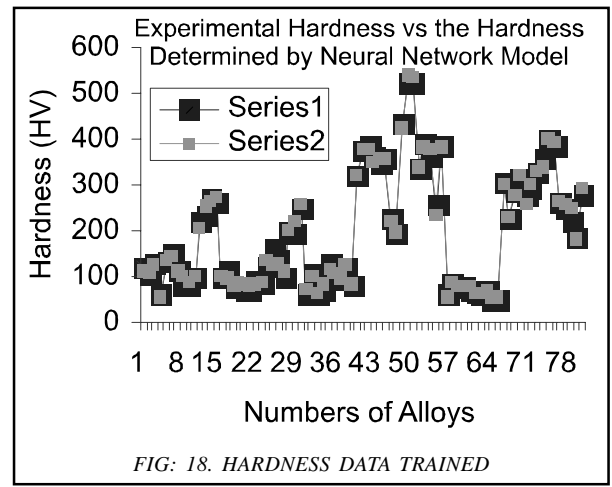

FIG: 18. HARDNESS DATA TRAINED

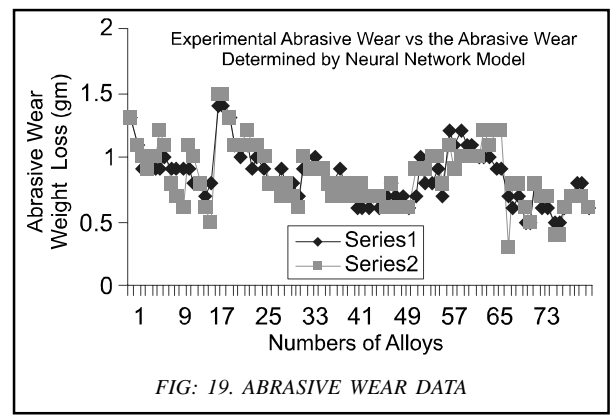

FIG: 19. ABRASIVE WEAR DATA

Recently another work was conducted by Xing, et. al. [20]. He used 25 datasets for training and five for validation in predicting the effect of heat treatment on hardness and abrasive wear of HVHSS steels. Similarly Ozerdam and Kolukisa [21] had predicted the mechanical properties of carbon steel bar. He used 33 samples for training and 11 samples for validation process.

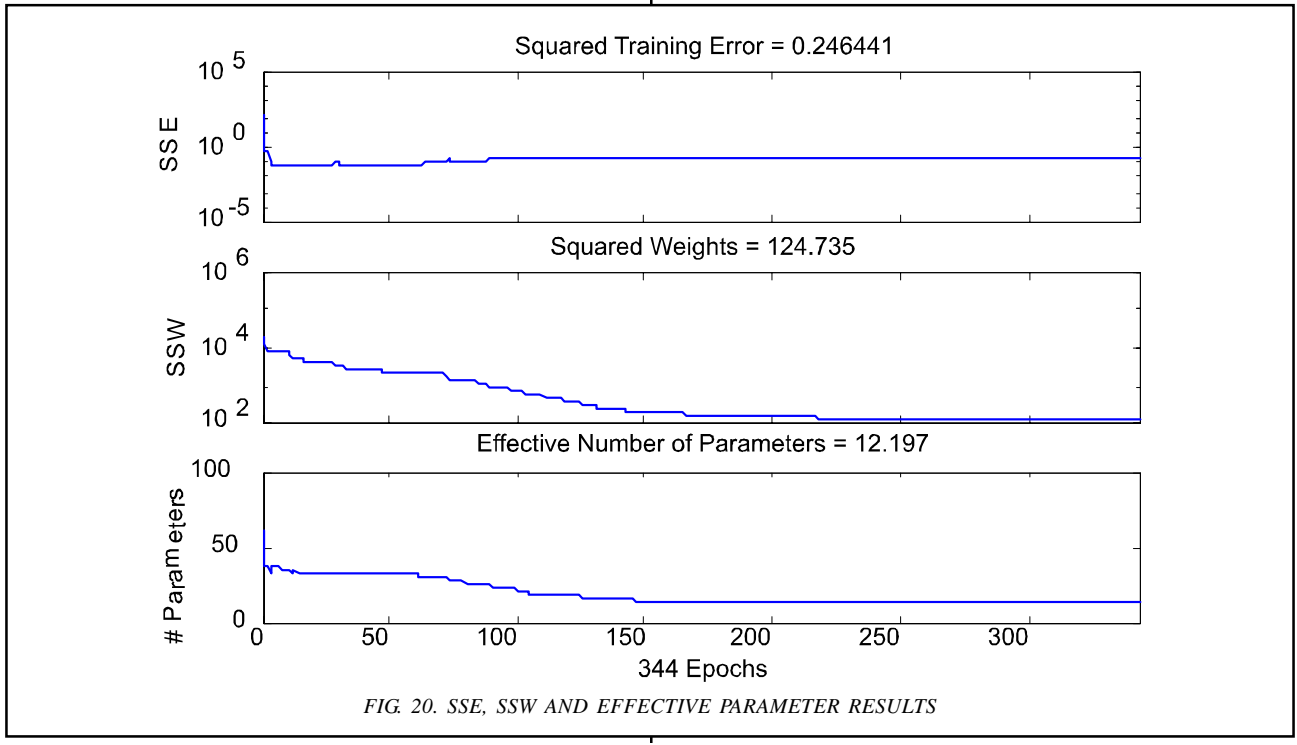

*FIG. 20. SSE, SSW AND EFFECTIVE PARAMETER RESULTS*

**TABLE 5. VALIDATION TABLE CONSISTS THE COMPARISON OF PROPERTIES OF AS CAST AS HEAT TREATED COMPOSITE DETERMINED BY LAB TESTS & ANNS MODEL FOR VALIDATION**

| No. | Alloy No. | Experimental Tensile Strength (MPa) | NN Tensile Strength (MPa) | Experimental Elongation (%) | NN Elongation (%) | Experimental HV | Neural Network HV | Experimental Weight Loss (gm) | NN Weight Loss (gm) |
|---|---|---|---|---|---|---|---|---|---|
| 1. | 1 2.5C | 160 | 155.9 | 16.00 | 13 | 130 | 139.9 | 0.9 | 1.0 |
| 2. | 3.5C | 100 | 96.9 | 12.00 | 9.3 | 240 | 250 | 0.858 | 0.8 |
| 3. | 3 15CB | 106.25 | 110 | 9.77 | 9.4 | 270 | 264 | 0.6 | 0.8 |
| 4. | 5 2.5C | 156.22 | 165 | 10.66 | 13 | 123 | 106 | 1.0 | 0.9 |
| 5. | 6 10C | 105.68 | 115 | 6.060 | 10 | 200 | 191 | 0.6 | 0.8 |
| 6. | 1 2.5HT | 186.57 | 195.3 | 15.49 | 19 | 140 | 157 | 0.89 | 0.9 |
| 7. | 3 2.5EHT | 165 | 166.9 | 14.00 | 11.7 | 430 | 457 | 0.63 | 0.7 |
| 8. | 4 2.5B | 161.97 | 160.8 | 16.18 | 17 | 80 | 79.2 | 0.92 | 1.2 |
| 9. | 4 10HT | 144.5 | 145.8 | 14.7- | 17 | 115 | 124 | 1.02 | 1.0 |
| 10. | 5 3.5 HT | 222 | 229.8 | 10.75 | 13.5 | 340 | 377 | 0.6 | 0.5 |

As a result of discussion above clearly the present investigation has more optimum approach for training as well as validation of the model because it contain 80 datasets for training and 10 datasets for validation. A close relation between training and validation was archived from the modeling results are shown in Figs. 25-26. The resulting error is 0.7 which was due to scattering of the database such as minimum and maximum variation in Si and Fe content in samples.

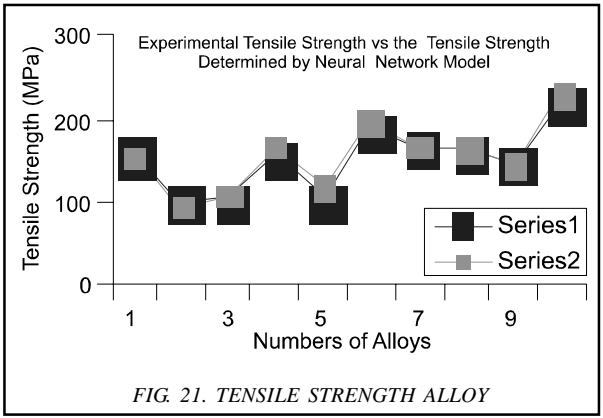

FIG. 21. TENSILE STRENGTH ALLOY

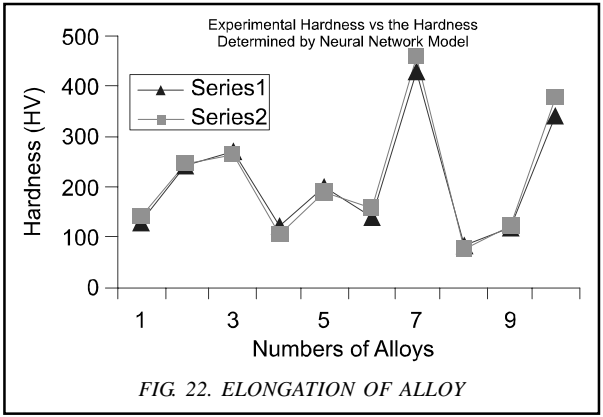

FIG. 22. ELONGATION OF ALLOY

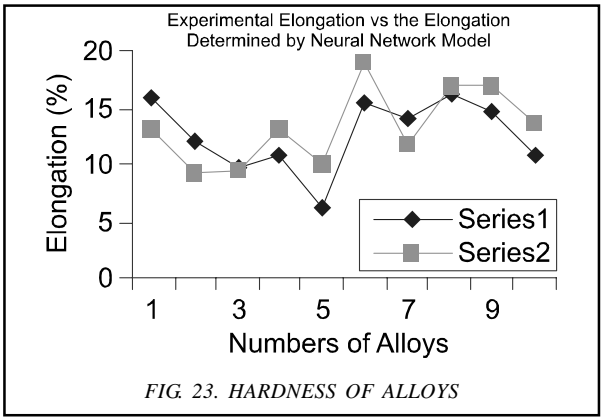

FIG. 23. HARDNESS OF ALLOYS

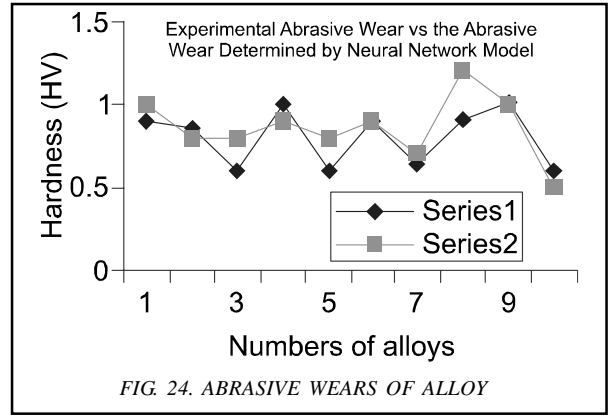

FIG. 24. ABRASIVE WEARS OF ALLOY

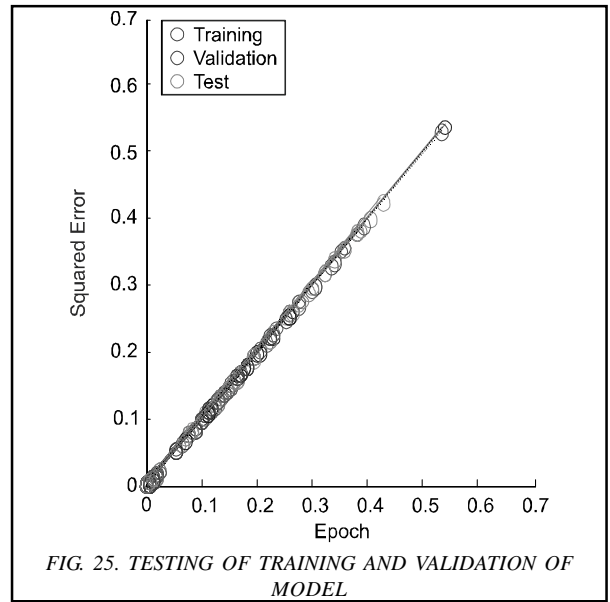

FIG. 25. TESTING OF TRAINING AND VALIDATION OF MODEL

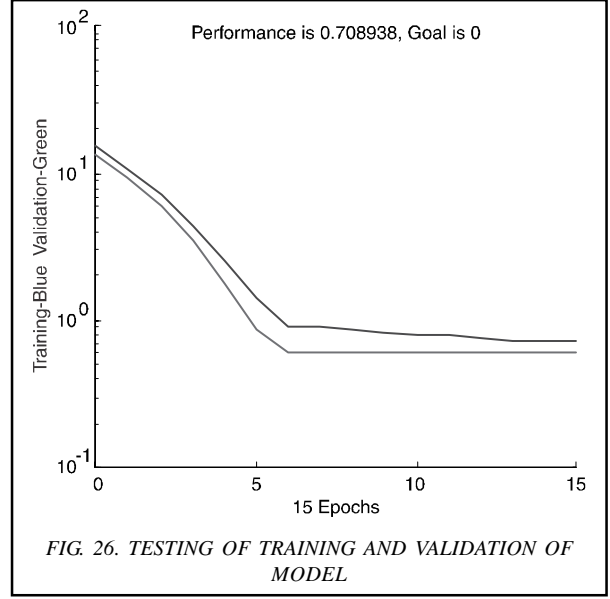

FIG. 26. TESTING OF TRAINING AND VALIDATION OF MODEL

## 4.  CONCLUSIONS

(i)  Mechanical properties of aluminum cast composite materials can successfully be modeled using artificial neural network which is an initial attempt to correlate the non linear behaviour between chemical composition and properties of cast aluminum composite developed via conventional foundry method carried out at Mehran University of Engineering and Technology, Jamshoro.

(ii)  A more comprehensive model was successfully developed with 80 dataset considering nonlinear relationship between the composition, processing parameters, and solution treatment with tensile strength, elongation, hardness and abrasive wear resistance using multilayer perceptron network.

(iii)  A well trained model with 9 hidden neurons giving smaller training error 2-7% and has better performance as compared to lesser or higher number of neurons  The suggested network model contain 14 inputs, 9 hidden neurons with 4 outputs. The model can be trained within 30 second having good generalization ability. Present proposed model predicts accurately the output of the unseen test data.

(iv)  The architecture of the model is MLP with back propagation learning algorithm. This model after successful training can more effectively be used not only for predicting various mechanical properties such as tensile strength, ductility (Elongation), hardness and abrasive wear resistance of aluminum cast composites reinforced with $Al_2O_3$ particles but can be used in design and development of aluminum cast composite materials.

(v)  ANN based model showing good agreement with the experimental results including training, testing (validation). By using present model for composite material design, development, can reduce large experimental work. The applications of proposed model not only reduce the overall experimental fatigue but also reduce the cost by optimizing the composition and process parameters.

## ACKNOWLEDGEMENTS

Authors acknowledge the help and support provided by the authority of Mehran University of Engineering & Technology, Jamshoro, Pakistan, to provide material and facility for research work. Thanks also extended to Professor Dr. Muhamad Moazam Baloch, and Dr. Muhammad Ishaque Abro, for their cooperation in conducting experimental work and Professor Dr. Naseem, Ex-Principal, Dawood College of Engineering & Technology, Karachi, Pakistan, to provide lab facility for tensile test.